# Pore size estimation in axon-mimicking microfibres with diffusion-relaxation MRI

Erick J. Canales-Rodríguez[1,2,*], Marco Pizzolato[2,3], Feng-Lei Zhou[4,5], Muhamed Barakovic[6], Jean-Philippe Thiran[1,7,8], Derek K. Jones[9], Geoffrey J.M. Parker[4,10,11], Tim B. Dyrby[2,3].

[1] Signal Processing Laboratory 5 (LTS5), Ecole Polytechnique Fédérale de Lausanne (EPFL), Lausanne, Switzerland.

[2] Danish Research Centre for Magnetic Resonance (DRCMR), Centre for Functional and Diagnostic Imaging and Research, Copenhagen University Hospital Amager and Hvidovre, Copenhagen, Denmark.

[3] Department of Applied Mathematics and Computer Science, Technical University of Denmark, Kongens, Lyngby, Denmark.

[4] Centre for Medical Image Computing, Department of Medical Physics and Biomedical Engineering, University College London, London, United Kingdom

[5] MicroPhantoms Limited, Cambridge, United Kingdom

[6] Translational Imaging in Neurology (ThINk) Basel, Department of Biomedical Engineering, University Hospital Basel and University of Basel, Basel, Switzerland.

[7] Radiology Department, Centre Hospitalier Universitaire Vaudois and University of Lausanne, Lausanne, Switzerland.

[8] Centre d'Imagerie Biomédicale (CIBM), EPFL, Lausanne, Switzerland.

[9] Cardiff University Brain Research Imaging Centre, Cardiff University, Cardiff, Wales, United Kingdom.

[10] Department of Neuroinflammation, Queen Square Institute of Neurology, University College London, London, United Kingdom

[11] Bioxydyn Limited, Manchester, United Kingdom

*Corresponding author (Erick J. Canales-Rodríguez), email: erick.canalesrodriguez@epfl.ch

Signal Processing Laboratory (LTS5), EPFL-STI-IEL-LTS5, Station 11, CH-1015 Lausanne, Switzerland

Word count: 6716 (including the body of the text, appendices, data availability statement, acknowledgements, and declaration of interest).

**Abstract** (247 words)

**Purpose**: This study aims to evaluate two distinct approaches for fibre radius estimation using diffusion-relaxation MRI data acquired in biomimetic microfibre phantoms that mimic hollow axons. The methods considered are the spherical mean power-law approach and a $T_2$-based pore size estimation technique.

**Theory and Methods**: A general diffusion-relaxation theoretical model for the spherical mean signal from water molecules within a distribution of cylinders with varying radii was introduced, encompassing the evaluated models as particular cases. Additionally, a new numerical approach was presented for estimating effective radii (i.e., MRI-visible mean radii) from the ground truth radii distributions, not reliant on previous theoretical approximations and adaptable to various acquisition sequences. The ground truth radii were obtained from Scanning Electron Microscope images.

**Results**: Both methods show a linear relationship between effective radii estimated from MRI data and ground-truth radii distributions, though some discrepancies were observed. The spherical mean power-law method overestimated fibre radii. Conversely, the $T_2$-based method exhibited higher sensitivity to smaller fibre radii but faced limitations in accurately estimating the radius in one particular phantom, possibly due to material-specific relaxation changes.

**Conclusion**: The study demonstrates the feasibility of both techniques to predict pore sizes of hollow microfibres. The $T_2$-based technique, unlike the spherical mean power-law method, does not demand ultra-high diffusion gradients but requires calibration with known radius distributions. This research contributes to the ongoing development and evaluation of neuroimaging techniques for fibre radius estimation, highlights the advantages and limitations of both methods and provides datasets for reproducible research.

# 1. Introduction

Accurately measuring the diameter of axons in vivo has been a crucial goal in diffusion MRI (dMRI) (1,2,11–13,3–10), as the axon diameter modulates the speed of action potentials along the axon and may serve as a biomarker of axonal degeneration (14–17). However, existing dMRI techniques are affected by a resolution limit, or diameter lower-bound, below which smaller axons cannot be detected. This limit is determined by the experimental setup, particularly the strength of the diffusion encoding gradient and the signal-to-noise ratio (SNR) (4,18). Unfortunately, dMRI signals collected in 3T clinical scanners equipped with diffusion gradients below 80 mT/m have a higher resolution limit (18), allowing only the detection of large axons. For additional discussions, the reader is referred to (19–24). As a result, dMRI-based diameter estimation techniques are primarily implemented in advanced human 'Connectom' scanners with stronger diffusion gradients (i.e., 300 mT/m) (25) and preclinical scanners (4). A recent study has shown that the effective radius – defined as the 'apparent' MRI-visible mean axon radius representing the entire axon radius distribution within a voxel – can be estimated by eliminating two crucial confounding factors from the dMRI signal that affected previous studies: extra-axonal water and axonal orientation dispersion (2). This method is referred to as the spherical mean power-law approach.

Alternatively, in porous media and tissues, pore and cell sizes can be estimated using a surface-based $T_2$ relaxation model (26–31). This model predicts a linear dependence between the inverse of the intra-pore/cell $T_2$ and the surface-to-volume ratio of the confining pore/cell geometry (32), which is proportional to the inverse of the radius for a cylinder. However, this technique cannot be directly applied in living tissue to estimate axon radius since the $T_2$ measured by conventional quantitative MRI techniques is affected by both the intra-axonal and extra-axonal water compartments. To overcome this limitation, we recently proposed a new diffusion-relaxation MRI approach for quantifying axon radii (33,34) based on estimating the intra-axonal $T_2$ relaxation time (35).

In practice, both approaches for estimating axon radii – the spherical mean power-law method (2) and the $T_2$-based pore size estimation technique (34) – involve collecting a first dMRI dataset using a fixed echo time (*TE*>50 ms) long enough to attenuate the myelin water dMRI signal (36), and multiple diffusion gradients orientations with a high *b*-value (e.g., *b*≥4000-6000 s/mm$^2$ for in vivo data) to attenuate the extra-axonal dMRI signal (37). This way, the acquisition parameters act as a filter, significantly reducing the contribution of water molecules from all white matter compartments to the measured dMRI signal but not the intra-axonal space. The difference between the two acquisition approaches lies in how the second data block is acquired. The spherical mean power-law method requires measuring another dMRI dataset using the same *TE* and much higher *b*-values (e.g., $b \geq 10000$ s/mm$^2$ for in vivo data) with ultra-strong diffusion gradients only available on specific scanners. This is needed to reduce the resolution limit (18). On the other hand, the T$_2$-based pore size estimation method requires collecting another dMRI dataset using the same *b*-value employed in the first acquisition block but using different *TE*s. In both techniques, the dimensionality of the data is reduced before fitting the models by computing the orientation-averaged spherical mean signal, which is a rotationally invariant metric that does not depend on the underlying fibre orientation distribution (38). This strategy effectively reduces the number of parameters estimated in the diffusion and relaxation models.

In the spherical mean power-law approach (2), the intra-axonal radial diffusivity $D_\perp$ is calculated from the dMRI data acquired with high and ultra-high *b*-values, which is then converted into a radius by using the van Gelderen model based on the Gaussian phase distribution approximation (39). Conversely, for the T$_2$-based pore size estimation method (34), the intra-axonal T$_2$ time is determined from the dMRI data acquired using multiple *TE*s, following the approach suggested by (35). The intra-axonal T$_2$ is subsequently converted into a radius using a surface-based relaxation model (31,40) that requires a calibration process to determine the T$_2$ surface relaxivity, an unknown parameter that depends on the relaxation properties of the inner axon surface (34). The main practical disadvantage of the calibration step is that it requires knowing the ground-truth radius in some brain regions, information that is not always available.

Although both methods hold great promise for accurate pore/cell size estimation, a systematic evaluation of these techniques in a controlled setting with a known ground truth has yet to be conducted. Additionally, the absence of a comprehensive multi-contrast diffusion-relaxation model for the spherical mean signal generated by water molecules within a distribution of pore sizes (or axon radii) represents a notable limitation that hinders our ability to elucidate the theoretical relationship between these techniques.

To overcome these limitations, this study outlines the following objectives: (1) Formulate a theoretical diffusion-relaxation model capable of encompassing both the spherical mean power-law and T2-based methods. This formulation shall be helpful to clarify the main assumptions underpinning each approach. (2) Evaluate both techniques using diffusion-relaxation MRI data acquired in biomimetic phantoms where the ground truth is known. These phantoms consist of co-electrospun hollow axon-mimicking microfibres with non-circular cross-sections and different radii distributions. (3) Introduce a novel numerical approach for calculating effective radius from radius distributions measured via scanning electron microscopy (SEM). This numerical approach is necessary to circumvent the limitations associated with previous approximated analytical expressions, which are not accurate for the range of pore sizes in the employed phantoms.

## 2. Theory

*2.1 Intra-pore diffusion-relaxation MRI model*

This section introduces a diffusion-relaxation model for the spherical mean MRI signal generated by water molecules filling the intra-pore space of a distribution of cylinders with different radii. This formulation unifies into a single model the two techniques evaluated in this study:

$$\bar{S}(b,TE) = k \frac{\int P(r) r^2 \bar{S}_{\text{Rel}}(TE,r) \bar{S}_{\text{Diff}}(b,r) dr}{\int P(r) r^2 dr} \quad , \qquad [1]$$

where the spherical mean diffusion-relaxation signal $\bar{S}(b,TE)$ depends on the *b*-value and echo time (*TE*), $k$ is a constant proportional to the total intra-pore volume, $r$ denotes the radius, $P(r)$ is the radius distribution, and the volumetric correction factor $r^2$ accounts for the volume-weighted nature of the measured MRI signal (i.e., the signal intensity from each cylinder is proportional to the number of water molecules inside the cylinder, and thus, to its volume).

The T2 relaxation-weighted MRI signal $\bar{S}_{\text{Rel}}(TE,r)$ for a cylinder with radius $r$ is

$$\bar{S}_{\text{Rel}}(TE,r) = \exp\left(-\frac{TE}{T_2^i(r)}\right), \qquad [2]$$

where the intra-pore transversal relaxation time $T_2^i$ depends on $r$, according to (32,34,40)

$$\frac{1}{T_2^i} = \frac{1}{T_2^b} + \frac{2\rho_2}{r} \ , \quad [3]$$

where $T_2^b$ denotes the T$_2$ relaxation time of the bulk (free) water filling the cylinders and $\rho_2$ is the T$_2$ surface relaxivity depending on the phantom material.

The spherical mean diffusion-weighted signal $\bar{S}_{\text{Diff}}(b,r)$ from a cylinder with radius $r$, in Eq. [1], is modelled as

$$\bar{S}_{\text{Diff}}(b,r) = \sqrt{\frac{\pi}{4}} \exp(-D_\perp(r)b) \frac{erf\left(\sqrt{b(D_\parallel - D_\perp(r))}\right)}{\sqrt{b(D_\parallel - D_\perp(r))}} \ , \quad [4]$$

which is the spherical mean signal equation for an axis-symmetric diffusion tensor (2,10,37,38,41,42), where $erf$ denotes the error function, and the radial diffusivity $D_\perp$ depends on $r$ according to the van Gelderen model (39), defined in Eq. [10] in Appendix A.

In the following two subsections, we will examine the necessary approximations required to derive the T$_2$-based pore size estimation technique (34) and the spherical mean power-law method (2) from the more general model presented in Eqs. (1)-[4].

*2.2 Intra-pore pure relaxation MRI model: T$_2$-based estimation technique*

When the data is measured using a diffusion gradient that is not sufficiently strong, the sensitivity of the diffusion-weighted signal to the cylinder radius is significantly reduced (18), i.e., the diffusion signal becomes proportional to the signal from a cylinder with infinitesimal radius, $\bar{S}_{\text{Diff}}(b,r) \propto \bar{S}_{\text{Diff}}(b,r \to 0)$. In such cases, $\bar{S}_{\text{Diff}}(b,r)$ can be treated as a constant and moved outside the integral in Eq. [1]. As a result, the general diffusion-relaxation model becomes a pure relaxation model:

$$\bar{S}(TE) = K \frac{\int P(r) r^2 \bar{S}_{\text{Rel}}(TE, r) dr}{\int P(r) r^2 dr} \quad , \qquad [5]$$

where $K$ is a constant to be estimated that absorbed the diffusion signal. Similar to the spherical mean power-law method described in the next subsection, the integral in Eq. [5] is approximated by the relaxation signal from a single cylinder with an effective radius characterising the whole distribution

$$\bar{S}(TE) \approx K \bar{S}_{\text{Rel}}(TE, r_{\text{eff}-MRI-R}) . \qquad [6]$$

Notably, the T$_2$-based pore size estimation technique, as proposed in (34), relies on the pure relaxation model defined in Eq. [6]. Note that $r_{\text{eff}-MRI-R}$ denotes the effective MRI-visible radius resulting from the relaxation process.

*2.3 Intra-pore pure diffusion MRI model: spherical mean power-law*

If the relaxation signal $\bar{S}_{\text{Rel}}(TE, r)$ can be neglected in Eq. [1], e.g., by assuming that $T_2^i$ is a constant independent of $r$, when $\rho_2 \to 0$, then it can be treated as a constant term and moved outside the integral. Accordingly, the diffusion-relaxation model is simplified, resulting in a pure diffusion model

$$\bar{S}(b) \approx \beta \frac{\int P(r) r^2 \bar{S}_{\text{Diff}}(b, r) dr}{\int P(r) r^2 dr} , \qquad [7]$$

where $\beta$ is a constant to be estimated that absorbed the relaxation signal. The integral in Eq. [7] is approximated by the spherical mean diffusion signal from a cylinder with an effective radius:

$$\bar{S}(b) \approx \beta \bar{S}_{\text{Diff}}(b, r_{\text{eff}-MRI-D}) , \qquad [8]$$

where $r_{eff-MRI-D}$ denotes the effective MRI-visible radius resulting from the diffusion process. Notice that $r_{eff-MRI-D}$ and the effective radius calculated from the relaxation process described in the previous subsection $r_{eff-MRI-R}$ are not necessarily equal, as the MRI signals from both modalities may have different sensitivities to pore size.

It is important to note that the spherical mean power-law technique for estimating axon radius presented in (2) is based on Eq. [8]. In that study, however, the authors simplified the model by using two additional approximations: (1) the term involving the error function in Eq. [4] was omitted since it tends to one for the axon radii found in the brain, and (2) the van Gelderen model defined by Eq. [10] relating $D_\perp$ and $r$ was replaced by the wide pulse approximation derived by Neuman (43), which is valid for small radii and long pulses ($\Delta \gg \delta \gg r^2/D_\parallel$); for more details see Eq. [11] in Appendix A. However, since these approximations are not valid for large radii, such as those measured in our phantoms, in this study we estimated the axon radii employing the more general expressions given by Eqs. [8], [4], and [10] using the van Gelderen model, as suggested by (44).

For theoretical purposes only, in Appendix A Eq. [12] (see also **Figure A1** of Appendix C), we introduce a new approximation for a broader application in the regime of medium-pulse times, $\Delta \gg \delta \gtrsim r^2/D_\parallel$, which is more accurate than Neuman's approximation for both small and large radii.

*2.4 Numerical effective radius*

We evaluate the relaxation and diffusion models in Eqs. [6] and [8] by comparing the effective radii $r_{eff-MRI-R}$ and $r_{eff-MRI-D}$ estimated from the MRI data with the actual effective radius of the biomimetic phantoms determined from the underlying radius distribution $P(r)$, which was measured in our study using scanning electron microscopy (SEM). However, the method for estimating the actual effective radius from $P(r)$ has a significant limitation. The standard formula for calculating the effective radius $r_{eff} \approx \left(\langle r^6 \rangle / \langle r^2 \rangle\right)^{1/4}$ from the 6[th] and 4[th] moments of $P(r)$ (2,45) was derived under the wide pulse approximation by Neuman. Therefore, it is only valid for small radii (44), much smaller than the ones measured in the phantoms. Hence, this formula cannot be used in the evaluation. For further information, refer to **Figure A1** in Appendix C.

To address this issue, we propose a new numerical approach to estimate effective radius from $P(r)$, which is valid for radius distributions with both small and large radii. This approach generates the synthetic relaxation $\bar{S}_{\text{Rel-SEM}}$ and diffusion $\bar{S}_{\text{Diff-SEM}}$ signals produced by the actual radius distribution $P(r)$. The synthetic signals are generated by discretising the integrals in Eqs. [5] and [7] using the measured radii, respectively:

$$\bar{S}_{\text{Rel-SEM}}(TE, \rho_2) \approx K \sum_{i=1}^{N} \left( \frac{r_i^2}{\sum_{j=1}^{N} r_j^2} \right) \bar{S}_{\text{Rel}}(TE, r_i, \rho_2),$$

$$\bar{S}_{\text{Diff-SEM}}(b) \approx \beta \sum_{i=1}^{N} \left( \frac{r_i^2}{\sum_{j=1}^{N} r_j^2} \right) \bar{S}_{\text{Diff}}(b, r_i),$$

[9]

where $\{r_i, i=1,...N\}$ denotes the set of $N$ radii measured per phantom.

By employing the same equations used to predict the effective radius from the MRI data, i.e., Eqs. [6] and [8], it is then possible to estimate the SEM-based effective radii $r_{\text{eff}-SEM-R}$ and $r_{\text{eff}-SEM-D}$ from these synthetic signals for the assumed relaxation and diffusion models.

Additionally, for the pure relaxation model, we consider a further approximation to estimate the effective radius by calculating the ratio of the second and first moments of $P(r)$, $r_{\text{eff}-SEM} = <r^2>/<r>$. This approximation is based on a Taylor expansion of the relaxation model, presented in Appendix B. Notice that $r_{\text{eff}-SEM}$ provides an approximation to the value of $r_{\text{eff}-SEM-R}$. It is estimated directly from the radius distribution and does not involve generating a synthetic relaxation signal.

3. **Methods**

*3.1 Phantom construction and characterisation*

Five phantom samples consisting of micron-scale hollow fibres mimicking axons in white matter were built using the co-electrospinning technique (46) to produce microfibres with a different distribution of inner fibre radius per phantom. Each phantom was constructed by concatenating various phantom samples (layer substrates) created to have similar distributions of fibre radii.

The inner fibre radii of each phantom were measured using five SEM images taken from different phantom samples. The SEM images were captured using a Phenom ProX desktop SEM (Thermo Fisher Scientific, USA) with an accelerating voltage of 5 kV. The ImageJ software (imagej.nih.gov/ij) was employed to analyse the SEM images, following the methods described in (47,48). Specifically, for each sample, the SEM images underwent a process of binarisation and thresholding. Subsequently, the intra-fibre area of each pore was determined utilising the 'Analyse Particles' feature in ImageJ. The automated measurements of each intra-fibre area ($A_i$) were then transformed into the corresponding inner fibre radius under the assumption that the cross-section of the pore is circular. This transformation was achieved through the formula $r_i = \sqrt{A_i/\pi}$, where $r_i$ represents the inner fibre radius of each ith measured pore. Therefore, the inner fibre radius is defined as the radius of a cylinder with the same cross-sectional area (or volume) as the pore, ensuring that the computed radius reflects the pore's volumetric properties.

Phantom 1, 3, 4, and 5 were composed of parallel fibres with different radii, while Phantom 2 comprised two groups of parallel fibres with an inter-fibre angle of 90°. Phantom 1 and Phantom 2 were designed to have similar distributions of fibre radius.

All phantoms were placed inside 15 mL centrifuge tubes filled with de-ionised water. An additional control tube only containing de-ionised free water was also studied. The control tube was used to estimate the diffusion coefficient and $T_2$ relaxation time of the de-ionised water.

*3.2 Data acquisition*

Diffusion-relaxation and multi-shell dMRI data were collected using a 7T Bruker preclinical scanner at the Danish Research Center for Magnetic Resonance (DRCMR). Airflow at a controlled room temperature was applied around the sample to ensure a steady sample temperature during the acquisition. The diffusion-relaxation protocol used to fit the pure relaxation model (i.e., $T_2$-based technique) had the following acquisition parameters: a *b*-value of 5000 s/mm$^2$ (diffusion gradient, *G*=166.8 mT/m; diffusion times, $\Delta/\delta$=35/9 ms) acquired in 48 equidistant diffusion directions distributed over the unit sphere; a repetition time *TR* of 6100 ms; a voxel-size of 2x2x2 mm$^3$; and one *b*=0 s/mm$^2$ image per echo time (*TE*). The acquisition was repeated for six *TE*s: [51, 75, 100, 150, 200, 250] ms.

The multi-shell dMRI acquisition protocol employed to fit the pure diffusion model (i.e., spherical mean power-law) consisted of using five high *b*-values: *b*=[5000, 6000, 7000, 8000, 10000] s/mm$^2$ with respective diffusion gradients *G*=[166.8, 182.7, 197.3, 210.95, 235.85] mT/m. The *TE* was set to 51 ms, and one *b*=0 s/mm$^2$ (b0) image was acquired per *b*-value. The other experimental parameters, such as *TR*, $\Delta$, $\delta$, voxel size, and the number of diffusion directions, were kept the same as in the diffusion-relaxation acquisition sequence. To evaluate the SNR in our experiments, we employed the following approach. For each voxel, we calculated the SNR as the ratio of the mean value to the standard deviation across the set of five b0 images. Subsequently, the mean SNR was determined from the individual voxelwise SNR values within a mask comprising the five phantoms, yielding a mean SNR value of 34. In our experiments, we opted to utilise the raw data without preprocessing because the phantom data remained unaffected by motion. Introducing any denoising step was avoided to prevent unwanted smoothing effects and partial volume contamination, especially given the small size of the phantoms.

*3.3 Estimation*

Like in previous studies (35,49–51), we assumed that for *b*≥5000 s/mm$^2$ the signals originating from water molecules outside the intra-fibre compartment, which likely experience larger diffusion displacements, are highly attenuated. This assumption allows us to focus on the signals originating within the intra-fibre compartment. We computed the spherical mean signal $\overline{S}(TE,b)$ by averaging the signal measurements over all the diffusion gradient directions (38,52–54) for each *b*-value and *TE*.

From the diffusion-relaxation data acquired at different *TE*s, we estimated the relaxation time within the intra-fibre compartment $T_2^i$ by fitting the mono-exponential relaxation model (35,50,51) defined by Eqs. [6] and [2]. To perform the fitting, we employed the non-linear 'L-BFGS-B' optimisation method available in Scipy (55). Subsequently, we implemented a calibration approach to estimate $\rho_2$, which enables us to calculate $r_{eff-MRI-R}$ from the intra-fibre $T_2^i$ times using Eq. [3]. The following subsection provides additional information on the calibration procedure.

To obtain the effective radius from the pure diffusion model $r_{eff-MRI-D}$, i.e., spherical mean power-law method, we fitted Eqs. [8], [4] and [10] to the multi-shell dMRI data. In Eq. [10], we included the first $m=18$ terms in the series to capture the diffusion behaviour within the fibres accurately (2).

The estimated radii $r_{eff-MRI-R}$ and $r_{eff-MRI-D}$ were compared with the SEM-based effective radii $r_{eff-SEM-R}$ and $r_{eff-SEM-D}$ derived from the underlying radius distributions $P(r)$, respectively. The effective radii $r_{eff-SEM-R}$ and $r_{eff-SEM-D}$ were calculated as described in the *Numerical effective radius* subsection using custom in-house software.

### 3.4 $T_2$-based calibration to estimate the surface relaxivity

To predict the fibre radius $r_{eff-MRI-R}$ from the intra-fibre relaxation time $T_2^i$, it is necessary to determine the values of $T_2^b$ and $\rho_2$ in Eq. [3]. In porous media, the parameter $T_2^b$ is typically neglected as its value is significantly larger than $T_2^i$. In our study, we used the control tube to estimate it and found $T_2^b \approx 3s$. The surface relaxivity $\rho_2$ was calculated for each phantom by minimising the mean squared difference between the measured diffusion-relaxation data $\bar{S}(TE)$ and the synthetic relaxation signal $\bar{S}_{\text{Rel-SEM}}(TE, \rho_2)$ generated by Eq. [9] using the actual radius distribution measured by SEM.

Once these parameters ($T_2^b$ and $\rho_2$) are computed, the fibre radius can be estimated. Two approaches were considered in this study to evaluate the accuracy of the radius estimation: (1) assuming a constant surface relaxivity for all phantoms by calculating the mean ($\bar{\rho}_2$), and (2) using the individual optimal surface relaxivity value estimated for each phantom.

4. **Results**

*4.1 Electron microscopy analysis*

The SEM analysis was conducted to examine the morphology of the phantom fibres. **Figure 1** shows an example of SEM micrographs, visually representing the fibre structure. The inner fibre radius distribution for each phantom is depicted in **Figure 2**, allowing for a comprehensive understanding of the variations in fibre radii. While all phantoms exhibit a significant proportion of fibre radii that resemble those observed in human brains, it is important to note the presence of a notable population of larger radii ranging from 4-10 µm.

Insert Figure 1 here (1.5 columns)

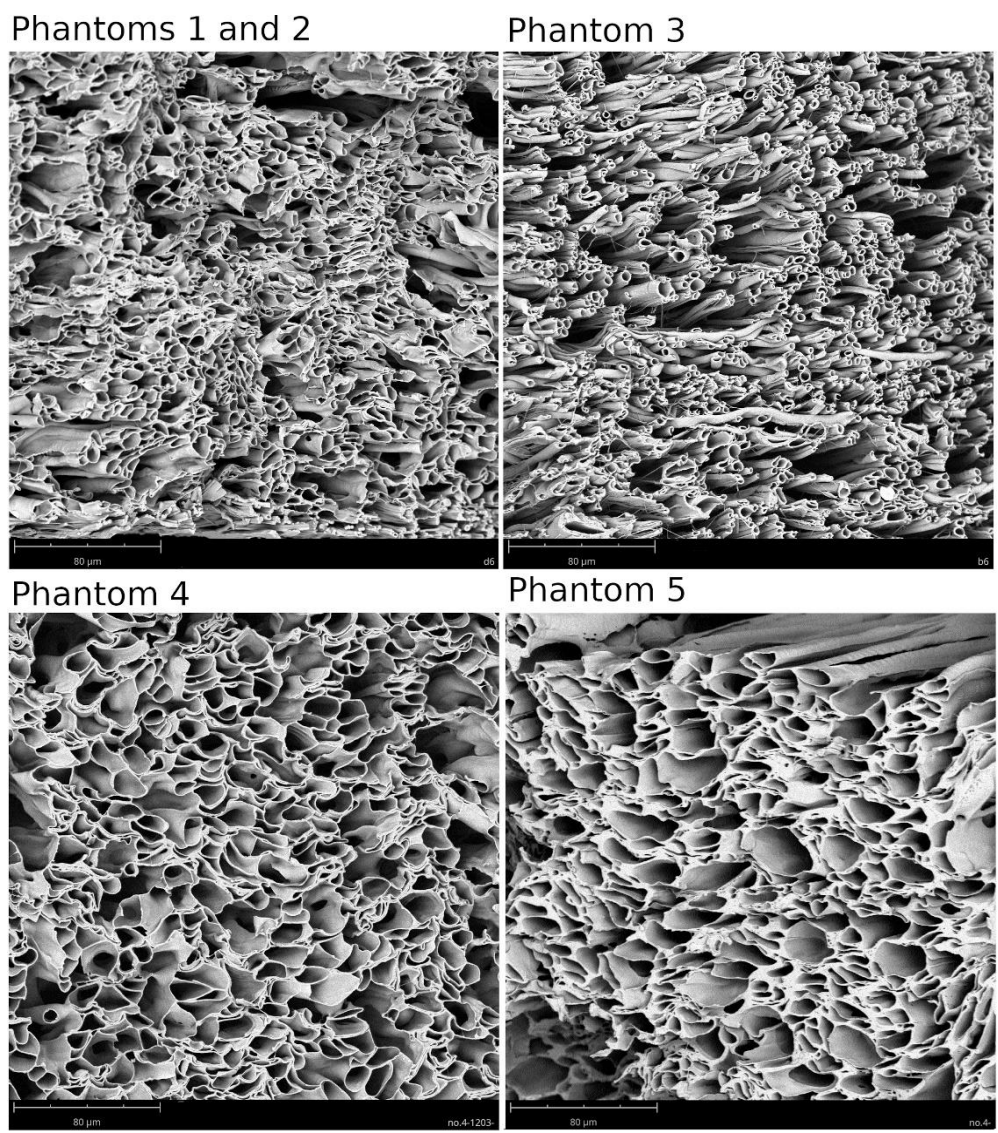

**Figure 1**. Scanning Electron Micrographs depicting the microscopic morphology of the biomimetic phantom samples. All phantom samples are presented on the same length scale (80 µm). Phantom 1 and Phantom 2 are shown together as they were constructed using similar distributions of fibre radii.

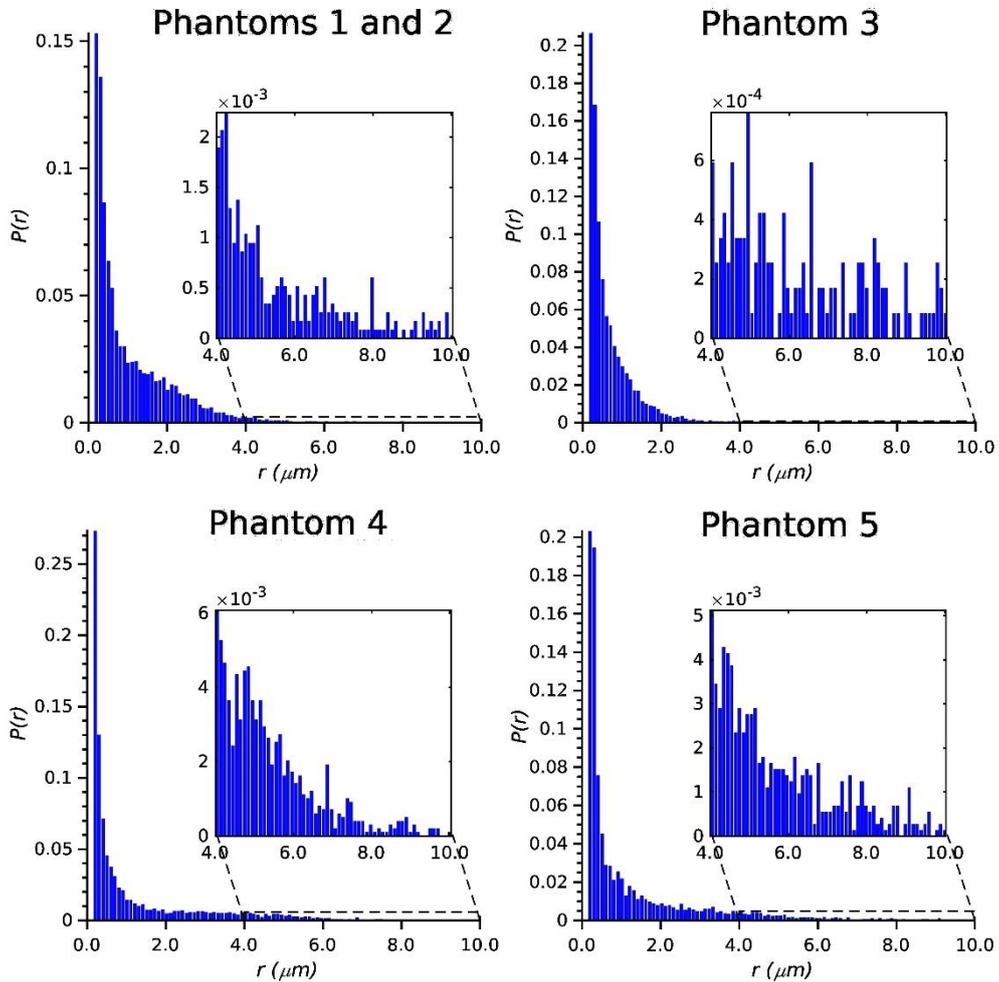

**Figure 2**. Radius distribution per phantom estimated using scanning electron microscopy. Phantom 1 and Phantom 2 are displayed together as both were built using similar distributions. The right tail of each distribution is zoomed in to visualise the distribution of the largest fibres. The mean radius and number of radii measured for each phantom are reported in Table 1.

**Table 1** provides quantitative data on the average fibre radius for each phantom and the sample size, i.e., the number of radii that were measured for the analysis.

Insert Table 1 here.

**Table 1**. Mean radius $\langle r \rangle$ per phantom, calculated from the radius distributions depicted in Fig 2. The number of measured radii $N$ using scanning electron microscopy is reported.

|  | $\langle r \rangle$ µm | $N$ |
|---|---|---|
| Phantoms 1&2 | 1.07 | 11618 |
| Phantom 3 | 0.70 | 11827 |
| Phantom 4 | 1.18 | 9880 |
| Phantom 5 | 1.21 | 7246 |

*4.2 T₂-based pore size estimation*

The T$_2$-based calibration analysis performed to estimate the surface relaxivity revealed that most phantoms exhibited a similar surface relaxivity value, with a mean of $\bar{\rho}_2 = 3.7 \pm 0.6$ nm/ms. However, Phantom 3 showed a reduced surface relaxivity, $\rho_2 = 2.0$ nm/ms, deviating from the average value observed in the other phantoms.

The comparison between the effective radii $r_{eff-MRI-R}$ calculated from the measured diffusion-relaxation data, assuming that all the phantoms have the same surface relaxivity $\bar{\rho}_2$, and the SEM-based effective radii $r_{eff-SEM-R}$ is presented in panel A of **Figure 3**. The estimates align closely with the 'y=x line of identity', indicating a nearly perfect linear relationship, except for Phantom 3, which substantially differs from this linear trend. The regression line fitted to the data has an intercept of 0.66 µm and a slope of 0.88.

The correlation coefficient ($R$) measuring the strength of the linear relationship between the two radii sets was not statistically significant, $R=0.66$, $P=0.228$. However, excluding Phantom 3 from the analysis made it statistically significant ($R=0.95$, $P=0.046$), indicating a strong linear relationship between the estimated radii for the remaining phantoms.

Panel B of **Figure 3** shows the measured MRI data and the generated synthetic signals from the ground truth radii distributions as a function of *TE*. Overall, there is a close agreement between the two data sets for all phantoms, except for Phantom 3, which exhibits notable discrepancies.

Insert Figure 3 (2 columns)

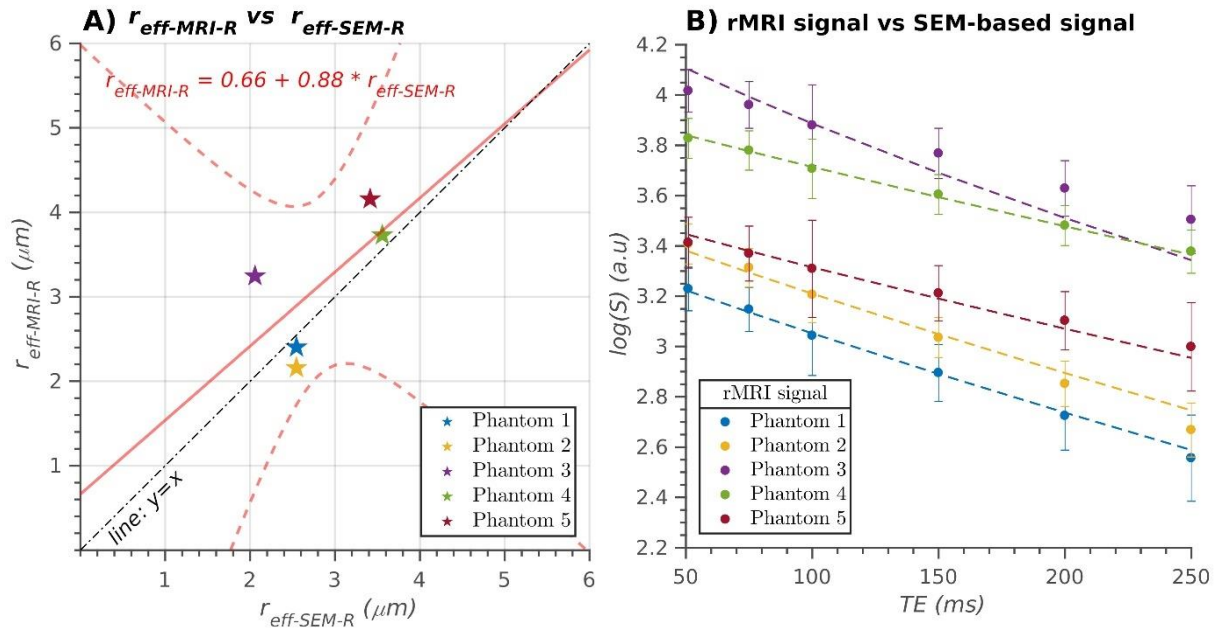

**Figure 3**. Panel A) presents the relationship between the $T_2$-based inner fibre radius (y-axis, $r_{eff-MRI-R}$) predicted using the measured diffusion-relaxation data with a fixed surface relaxivity of $\bar{\rho}_2$ =3.7 nm/ms and the effective radius estimated from the synthetic relaxation signal generated using the radius distribution obtained from scanning electron microscopy (SEM) images (x-axis, $r_{eff-SEM-R}$). The scatter plot represents the radius estimated from the mean signal for all voxels within each phantom. The regression line compares the estimates, while the reference line (y=x) indicates perfect linear agreement. Panel B) displays the logarithm of the measured relaxation data (rMRI) represented by the mean value and standard deviation across all voxels per phantom, along with the SEM-based generated synthetic signal as a function of the echo time (*TE*) in the whole interval.

The analysis considering a different surface relaxivity for each phantom is presented in **Figure 4**. Panel A displays the regression line comparing the effective radii, demonstrating a perfect agreement between the estimated radii and the radii derived from the SEM-measured distributions. The regression line has an intercept of -0.0046 µm and a slope of 1.001, indicating a nearly one-to-one correspondence between the two sets of radii. The correlation coefficient is 1.0, with a significant p-value of 7e-9, confirming the strong linear relationship. Panel B compares the measured diffusion-relaxation data and the synthetic signals generated from the respective SEM-based radii distributions. The two data sets show excellent agreement, with close correspondence across the entire range of *TE* values.

Insert Figure 4 (2 columns)

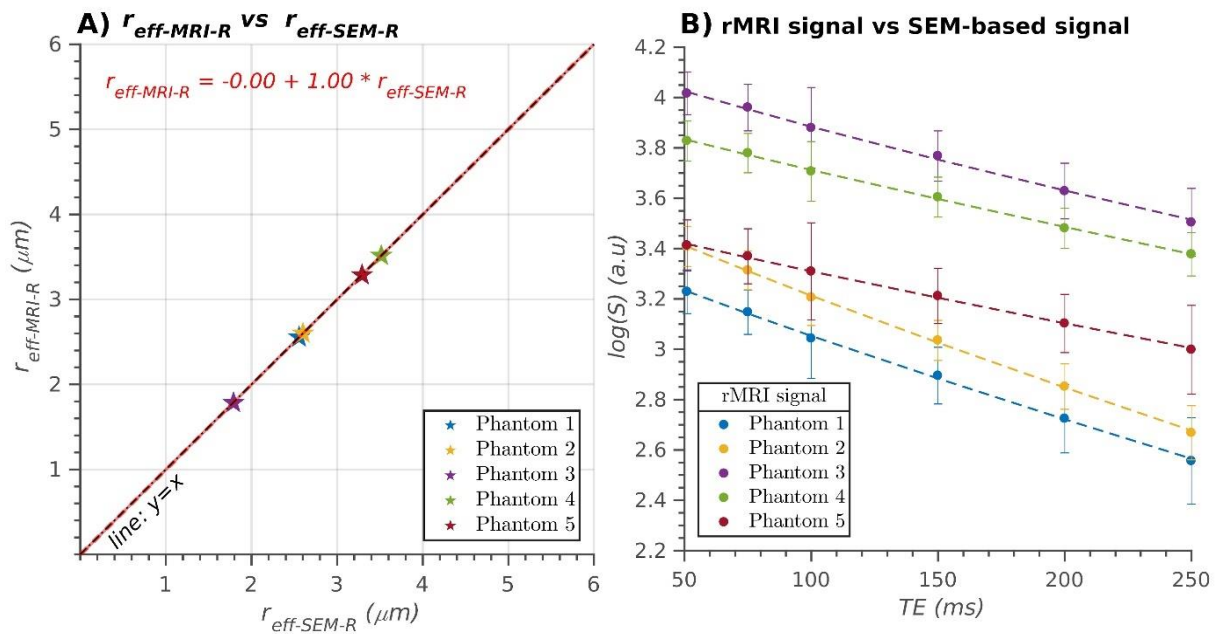

**Figure 4**. Panel A) presents the relationship between the $T_2$-based inner fibre radius (y-axis, $r_{eff\text{-}MRI\text{-}R}$) predicted using the measured diffusion-relaxation data with the surface relaxivity estimated individually for each phantom, and the effective radius calculated from the synthetic relaxation signal generated using the radius distribution obtained from scanning electron microscopy (SEM) images (x-axis, $r_{eff\text{-}SEM\text{-}R}$). The scatter plot represents the radius estimated from the mean signal for all voxels within each phantom. The regression line compares the estimates, while the reference line (y=x) indicates perfect linear agreement. Panel B) displays the logarithm of the measured relaxation data (rMRI) represented by the mean value and standard deviation across all voxels per phantom, along with the SEM-based generated synthetic signal as a function of the echo time (*TE*) in the whole interval.

In **Figure 5**, the linear relationship between the estimated effective radius $r_{eff-MRI-R}$ and the approximated effective radius derived from the second and first moments of the radius distribution, $r_{eff-SEM} = <r^2>/<r>$, is depicted. The regression analysis demonstrates a strong linear relationship between the radii, as indicated by the intercept of 0.15 µm and the slope of 0.93. The correlation coefficient (R=0.93) indicates a high degree of linear association between $r_{eff-MRI-R}$ and $r_{eff-SEM}$, which is statistically significant, *p*=0.0046.

Insert Figure 5 (1 column)

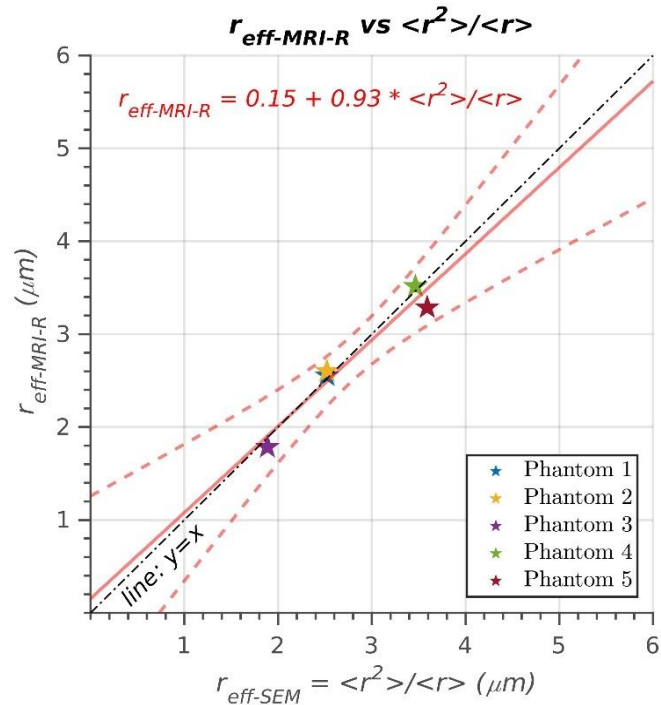

**Figure 5**. The linear relationship between the T$_2$-based inner fibre radius (y-axis, $r_{eff-MRI-R}$) predicted using the measured diffusion-relaxation data with the surface relaxivity estimated individually for each phantom (as in Figure 4) and the effective radius calculated from the moments of the radii distributions (x-axis, $r_{eff-SEM} = <r^2>/<r>$). The scatter plot represents the radius estimated from the mean signal across all voxels within each phantom. In addition to the regression line comparing both estimates, the reference line (y=x) is provided for visualising perfect linear agreement between the two measures.

*4.3 diffusion-based pore size estimation*

The diffusion-based pore size estimation analysis is presented in **Figure 6**. Panel A demonstrates the linear relationship between the fibre radii estimated from the multi-shell dMRI data using the spherical mean power-law approach and the radii derived from the SEM images. The statistically significant correlation coefficient (*R*=0.91, *P*=0.031) confirms the strength of this relationship. The intercept and slope of the regression line are 2.32 µm and 0.57, respectively. Panel B of **Figure 6** compares the measured multi-shell dMRI data and the synthetic diffusion signals generated from the SEM-based radius distribution. The plot shows the decay rates of the diffusion signals as a function of the *b*-value. It is observed that the synthetic diffusion signals exhibit lower decay rates compared to the measured data for all phantoms.

Insert Figure 6 here (2 columns)

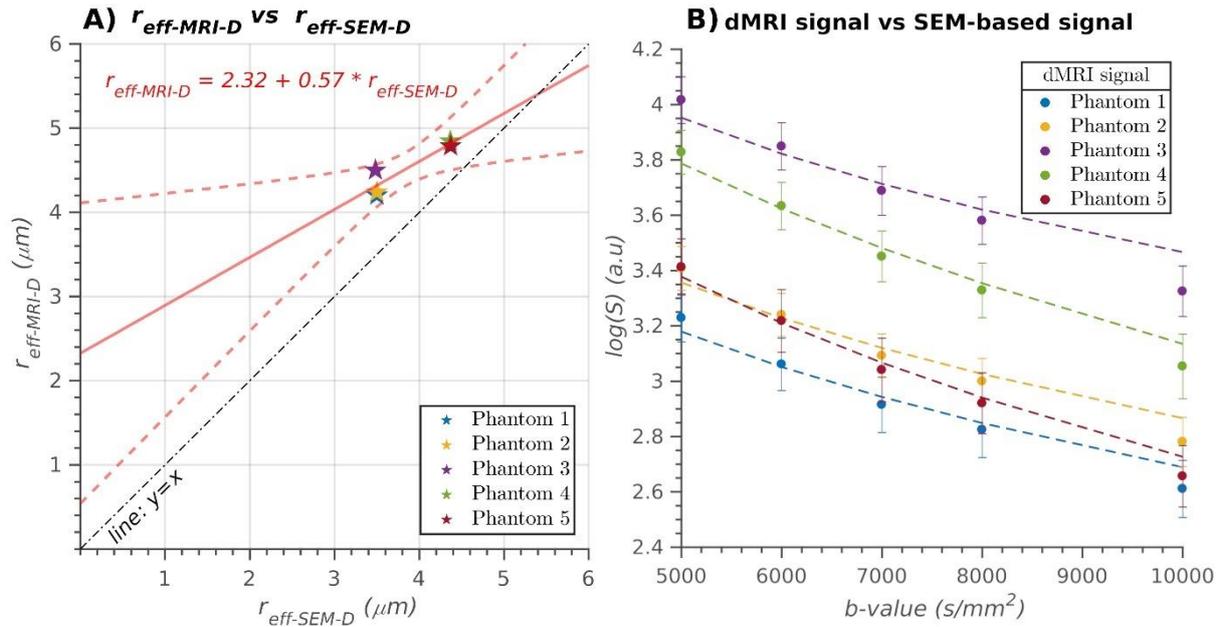

**Figure 6.** Panel A) illustrates the linear relationship between the dMRI-based fibre radius (y-axis, $r_{eff\text{-}MRI\text{-}D}$) estimated from the measured multi-shell dMRI data using the spherical mean power-law approach and the effective radius (x-axis, $r_{eff\text{-}SEM\text{-}D}$) calculated from the synthetic diffusion signal generated using the radii distributions measured from the scanning electron microscopy (SEM) images. Each data point represents the radius estimate obtained from the mean signal of all voxels within each phantom. The regression line compares the estimates from both methods, while the 'y=x line of identity' is a reference for perfect agreement. Panel B) depicts the logarithm of the measured multi-shell diffusion data (dMRI) and the SEM-based generated synthetic signal as a function of the *b*-value in the whole interval. The data points correspond to the mean values and standard deviations across all voxels within each phantom.

## 5. Discussion

We evaluated the spherical mean power-law method (2) and the $T_2$-based pore size estimation technique (34) using diffusion-relaxation MRI data acquired in biomimetic phantoms consisting of hollow axon-mimicking microfibres with non-circular cross-sections and different radii distributions. While the $T_2$-based pore size estimation technique requires a single high *b*-value and multiple (at least two) *TE*s, the spherical mean power-law method relies on a single *TE* and multiple (at least two) high *b*-values with very strong diffusion gradients. Notably, the $T_2$-based approach has more modest demands on the *b*-value than the diffusion-based spherical mean power-law technique. However, the $T_2$-based estimation approach relies on a calibration step that requires knowledge of the ground-truth radius distribution in specific regions to determine its surface relaxivity.

The linear relationship between the $T_2$-based effective radii estimated from the diffusion-relaxation MRI data and the ground truth radius distributions, as depicted in **Figures 3-5**, highlights the overall agreement between the estimates. However, it is worth noting that the estimation approach assuming a constant surface relaxivity for all phantoms was not accurate for Phantom 3, as evident from **Figure 3**, panel A. This deviation is attributed to the smaller surface relaxivity estimated for Phantom 3. Consequently, the predicted radius for Phantom 3 was considerably higher, leading to a mismatch between the generated synthetic signal and the measured data, as shown in panel B of **Figure 3**. The linear correlation coefficient for the estimated effective radii was not statistically significant. However, upon removing Phantom 3 from the analysis, the linear correlation coefficient became statistically significant, indicating a strong relationship between the estimated effective radii for the remaining phantoms. The reason behind the discrepancy in surface relaxivity for Phantom 3 remains uncertain. One plausible hypothesis is that, at the time of scanning, Phantom 3 underwent a natural degradation process typical of this type of material (56), resulting in altered interactions between water molecules and the pore surface. This hypothesis is further supported by the observation that, a few weeks after the MRI acquisitions, the white colour of Phantom 3 – unlike the other phantoms – turned to a light white-pink colour, indicating a change in its properties. Despite this issue, we decided to include the results of Phantom 3 in our study to provide a comprehensive analysis and present the complete findings.

The analysis using the individual surface relaxivity estimated for each phantom revealed a remarkable agreement between the effective radii, as demonstrated in **Figure 4**, panel A. Although this type of analysis is not practically feasible due to the requirement of knowing the radius distribution for each phantom, it serves as a valuable tool for model validation. Validating a model involves verifying whether the synthetic signal predicted by the model closely matches the measured data. While this criterion alone is insufficient to validate a model, as an incorrect over-parameterised model can still fit the data, it provides a necessary condition. In this study, the predicted synthetic signal for the relaxation model strongly agreed (**Figure 4**, panel B) with the measured data. To further explore the relationship between the $T_2$-based effective radius and the radius distribution, we conducted an additional analysis by replacing the effective radius used in **Figure 4**, estimated from the synthetic signals, with the effective radius calculated from the ratio of the second and first moments of the radius distribution. The results in **Figure 5** demonstrate that this relationship provides a good approximation.

On the other hand, the spherical mean power-law method exhibited a statistically significant linear relationship between the effective radii estimated from the multi-shell dMRI data and the ground truth radius distributions, as demonstrated in **Figure 6** (panel A), corroborating the sensitivity of this technique. However, the intercept of the linear regression line deviated considerably from zero, indicating an overestimation of the effective radius, particularly for phantoms with smaller radii. A closer examination of the synthetic signals generated by this model using the ground truth radius distributions (**Figure 6**, panel B) revealed notable discrepancies with the measured dMRI data. Specifically, the measured signals displayed a faster attenuation (i.e., a steeper slope of the logarithm of the signal as a function of the *b*-value), suggesting the presence of additional processes not accounted for in the model, which contributed to signal attenuation. One possible explanation for this discrepancy is the presence of numerical errors in accurately measuring the radius distribution used to generate the synthetic signals. However, while numerical errors cannot be entirely ruled out, they are unlikely to be the main contributor to the observed discrepancies because the SEM analysis measured several thousands of fibre radii per phantom (**Table 1**).

Upon comparing the effective radii obtained from the $T_2$-based and diffusion-based techniques, **Figures 3-5** and **Figure 6**, it is evident that these methods exhibit different sensitivities to spatial scales. Notably, the $T_2$-based approach demonstrates a higher sensitivity to smaller radii, resulting in smaller effective radii than the diffusion-based method. To further support this observation, we refer to **Figure A2** in Appendix C, where we present plots of the diffusion and relaxation signals as a function of the radius for the specific acquisition protocols employed in this study. Consistently with our findings, these plots highlight that the $T_2$-based method has a lower resolution limit for detecting small cylindrical fibres than the diffusion-based method. Nevertheless, the radii estimated by the spherical mean power-law method showed less variability.

Another contribution of this study is the numerical approach to estimate the effective radius from the underlying radius distribution. This approach offers several advantages as it does not rely on specific theoretical approximations. It can be applied more universally to different acquisition sequences, MRI contrasts, and materials with varying pore sizes. By comparing the generated synthetic signals with the measured data, this numerical approach allows us to assess the ability of the employed relaxation or diffusion models to explain the observed data. It is worth noting that previous studies proposed an expression for estimating the effective radius in dMRI based on the assumptions that the dMRI signal from the intra-fibre compartment can be approximated by the wide pulse or Neuman limit and that the diffusion model can be well-approximated by a first-order Taylor expansion, resulting in $r_{eff-SEM} = \left( \langle r^6 \rangle / \langle r^2 \rangle \right)^{1/4}$ (2,45). However, these assumptions only hold for a population of microfibres with radii smaller than 2.5 µm and do not apply to our study. In **Figure A1** of Appendix C, we conducted a supplementary analysis revealing that the signals derived from these approximations do not align with those predicted by the van Gelderen model across the entire range of measured radii in the phantoms. As a result, the new numerical approach proposed in this study becomes crucial for accurately determining the effective radius from the radius distribution.

Additionally, we introduced a general diffusion-relaxation theoretical model for the spherical mean signal originating from water molecules within a distribution of cylinders with varying radii. The two evaluated models are specific cases of this more comprehensive model. Examining the approximations made by each model provides valuable insights into their underlying assumptions. The pure-relaxation model provides a correct approximation for data acquired with high *b*-values, which effectively attenuates the extra-fibre signal. However, the diffusion gradients should not be strong enough to reduce the sensitivity of the data to the diffusion process inside the cylindrical pores. This setting may be more appropriate for clinical scanners with weaker diffusion gradients (~<100 mT/m). Conversely, the spherical mean power-law approach represents the solution to the general diffusion-relaxation model when the relaxation effect is neglected. In this case, it is less straightforward to determine how acquisition parameters should be adjusted to mitigate the influence of relaxation on the measured signal. Interestingly, by considering Eqs. [1], [7] and [8] it is possible to demonstrate that neglecting the relaxation effect in the spherical mean power-law approach leads to an effective radius estimate corresponding to a distorted radius distribution $\tilde{P}(r)$, which right-hand tail is inflated, leading to overestimated radii $\tilde{r}_{eff-MRI-D}$. This theoretical prediction aligns with the findings presented in **Figure 6**. For more technical details, see Appendix D.

It is important to mention that this is not the first study employing phantoms of hollow axon-mimicking fibres. Similar phantoms built with the co-electrospinning technique (46,57–59) have been used previously to validate other dMRI techniques, including diffusion tensor imaging and fibre tracking (46,60), microscopic fractional anisotropy using q-space trajectory encoding (61), anomalous diffusion (62), estimation of pore sizes in tumour tissue phantoms (63,64), as well as to investigate the stability and reproducibility of various dMRI-derived parameters (65), the validation of multidimensional dMRI sequences with spectrally modulated gradients (66), and to estimate pore sizes in similar complex microfibre environments using multi-shell dMRI (48).

This study has some limitations. First, the inner fibre radii estimated from SEM images are assumed to be the ground truth. However, the substrates generated per phantom are heterogeneous because it is not possible to control the resulting distributions of pore sizes in a precise way. As a result, different substrates from the same phantom had different distributions of pore sizes. To tackle this limitation, various SEM images from different substrates were used to estimate the mean effective radii per phantom. Accordingly, the effective radius predicted by the relaxation and diffusion models utilised the mean signal for all the voxels in the phantoms, and a voxelwise analysis was not possible. Second, the SEM-based radii were calculated by approximating intra-fibre pores as cylinders due to the inherent challenge of accurately representing irregular pore surfaces using MRI-based methods. This approximation, although essential, introduces potential biases. For example, we don´t know how the irregularity of the pore shape can deviate the measured dMRI data from the signal generated by a cylinder with the same volume (or cross-sectional surface area). Therefore, such potential discrepancies might have affected our results and could explain the signal differences observed in **Figure 6** panel B. However, it is worth noting that this issue is present in any clinical application of the evaluated methods. In brain data, there are other factors affecting the interpretation of results, including the effects of beading (radius variations along the axon), undulations (local variations in direction along the axon) and fibre dispersion (44,67,68). Addressing these limitations would require a more comprehensive technique capable of modelling these factors, which is beyond the scope of our work. Third, although the employed phantoms have a significant population of fibres with small radii, like those found in postmortem white matter axons, i.e., <1 µm (21,23,69,70), the proportion of fibres with larger sizes is much higher. Therefore, our findings should not be considered a strong demonstration of the validity of the employed techniques for estimating axon radius in brain white matter. Such a demonstration should require MRI data and histological analyses from the same brains. Fourth, as a single radial diffusivity and intra-fibre $T_2$ were estimated per phantom, the predicted radius is the effective radius. To determine the whole radius distribution, future studies should generalise the employed models to estimate distributions of diffusivities or $T_2$ times, respectively, e.g. see (71–78). Fifth, all our analyses employed raw diffusion-relaxation MRI data without preprocessing, so the Rician bias (79) may partially affect our results. Nevertheless, we verified that the SNR of our data was 34 and visually inspected the data to confirm our images were not dominated by noise. In a preliminary analysis (results not presented), we denoised the data using the MP-PCA (80) method and attenuated the Rician bias accordingly. However, we noted that the preprocessed data were slightly over-smoothed, and the correlation analysis comparing the estimated radii produced worse results. Therefore, we opted to employ the raw data to avoid the smoothing effects and prevent contamination of the diffusion-relaxation MRI signal by voxels outside the phantoms. Finally, despite our multi-shell dMRI acquisition protocol employed high and well-separated *b*-values (from 5000-10000 s/mm$^2$) to attenuate the dMRI signals from the extra-fibre pores significantly and to get 'enough' signal contrast to estimate the intra-fibre radial diffusivity, these *b*-values are not necessarily the optimal ones to assess the fibre radius. For example, in a previous study, *b*-values up to 30000 s/mm$^2$ were employed to estimate axon radii in the human white matter (81). Thus, our findings are specific to the implemented acquisition protocols and should not be extrapolated to other acquisition sequences and parameters.

6. Conclusions

This study demonstrates the feasibility of using intra-fibre $T_2$ times derived from diffusion-relaxation MRI data to predict the inner pore sizes of hollow axon-mimicking phantom fibres, as validated against measurements obtained from Scanning Electron Microscope images. Additionally, it confirms the sensitivity of the spherical mean power-law approach in estimating intra-fibre pore sizes from multi-shell dMRI data. The $T_2$-based estimation approach relies on a calibration step that requires knowledge of the ground-truth radius distribution in specific regions (phantoms) to determine its surface relaxivity. This limitation is absent in the pure dMRI model. However, the $T_2$-based estimation technique offers the advantage of using a smaller *b*-value. In contrast, the ultra-high diffusion gradients required by the dMRI-based approach are only achievable in preclinical or 'Connectom' 3T human scanners.

## 7. Data availability statement

The acquired MRI datasets and estimation scripts will be freely available at https://github.com/ejcanalesr/diffusion-relaxation-biomimetic-phantoms to facilitate reproducible research.

## 8. Acknowledgements

EJC-R is supported by the Swiss National Science Foundation (SNSF), Ambizione grant PZ00P2_185814. MP acknowledges funding from Danmarks Frie Forskningsfond (DFF) with case number 10.46540/3105-00129B. FLZ was supported by the NIHR UCLH Biomedical Research Centre (BRC) grant and UCL Department of Medical Physics and Biomedical Engineering and EPSRC (EP/M020533/1; CMIC Pump-Priming Award). This research was funded in whole, or in part, by a Wellcome Trust Investigator Award (096646/Z/11/Z) and a Wellcome Trust Strategic Award (104943/Z/14/Z). For the purpose of open access, the author has applied a CC BY public copyright licence to any Author Accepted Manuscript version arising from this submission. T.B.D. has received funding from the European Research Council (ERC) under the European Union's Horizon Europe research and innovation programme (grant agreement No.101044180). Views and opinions expressed are however those of the author(s) only and do not necessarily reflect those of the European Union or the European Research Council. Neither the European Union nor the granting authority can be held responsible for them.

## 9. Declaration of interest

Geoffrey Parker is a shareholder and director of Bioxydyn Limited, a company with an interest in imaging biomarkers. He is also a shareholder and director of Queen Square Analytics

## 10. Appendixes

*10.1 Appendix A*

The van Gelderen model (39), which is based on the Gaussian phase distribution approximation, relates the radial diffusivity $D_\perp$ and the radius $r$ as:

$$D_\perp = \frac{2\gamma^2 G^2 r^4}{bD_\|} \sum_{m=1}^{\infty} \frac{t_c}{\alpha_m^6 (\alpha_m^2 - 1)} \times \left[ 2\frac{\alpha_m^2}{t_c}\delta - 2 + 2e^{-\frac{\alpha_m^2 \delta}{t_c}} + 2e^{-\frac{\alpha_m^2 \Delta}{t_c}} - e^{-\frac{\alpha_m^2 (\Delta - \delta)}{t_c}} - e^{-\frac{\alpha_m^2 (\Delta + \delta)}{t_c}} \right], \quad [10]$$

where $D_\|$ is the intra-fibre parallel diffusivity, which is equal to the free diffusion coefficient when there is no restriction along the principal axes of the cylinders, $\gamma$ denotes the gyromagnetic ratio, $\delta/\Delta/G$ are the duration/separation/strength of the diffusion gradient, respectively, $t_c = r^2/D_\|$, $b = \gamma^2 G^2 \delta^2 (\Delta - \delta/3)$, and $\alpha_m$ are the roots of the derivative of the Bessel function of the first kind of order 1, $J_1'(\alpha_m) = 0$.

In the Neuman limit (43), i.e., $\Delta \gg \delta \gg r^2/D_\|$, Eq. [10] becomes

$$D_\perp(r) \approx \frac{7}{48} \frac{\gamma^2 G^2 \delta r^4}{bD_\|}. \quad [11]$$

In this work, we derived a new solution with a less restrictive limit. For $\Delta \gg \delta$ and $\delta \in (0, \sim r^2/D_\|)$, Eq. [10] becomes

$$D_\perp(r) \approx \frac{2\gamma^2 G^2 r^4}{bD_\parallel} \sum_{m=1}^{\infty} \frac{t_c}{\alpha_m^6(\alpha_m^2-1)} \left[ 2\alpha_m^2 \frac{\delta}{t_c} - 2 + 2e^{-\alpha_m^2 \frac{\delta}{t_c}} \right]$$

$$= \frac{4\gamma^2 G^2 r^4}{bD_\parallel} \sum_{m=1}^{\infty} \left[ \frac{\alpha_m^2 \delta}{\alpha_m^6(\alpha_m^2-1)} - \frac{t_c}{\alpha_m^6(\alpha_m^2-1)} + \frac{t_c}{\alpha_m^6(\alpha_m^2-1)} e^{-\alpha_m^2 \frac{\delta}{t_c}} \right]$$

$$= \frac{4\gamma^2 G^2 r^4}{bD_\parallel} \left[ \sum_{m=1}^{\infty} \frac{\alpha_m^2 \delta}{\alpha_m^6(\alpha_m^2-1)} - t_c \sum_{m=1}^{\infty} \frac{1}{\alpha_m^6(\alpha_m^2-1)} + t_c \sum_{m=1}^{\infty} \frac{1}{\alpha_m^6(\alpha_m^2-1)} e^{-\alpha_m^2 \frac{\delta}{t_c}} \right]$$

$$= \frac{4\gamma^2 G^2 r^4}{bD_\parallel} \left[ \delta \frac{7}{48 \times 4} - t_c \frac{7}{48 \times 4} \times \frac{12}{41} + t_c \frac{7}{48 \times 4} \times \frac{12}{41} e^{-\alpha_1^2 \frac{\delta}{t_c}} \right]$$

$$= \frac{4\gamma^2 G^2 r^4}{bD_\parallel} \left[ \delta \frac{7}{48 \times 4} - t_c \frac{7}{48 \times 4} \frac{12}{41} (1 - e^{-\alpha_1^2 \frac{\delta}{t_c}}) \right]$$

$$= \frac{7}{48} \frac{\gamma^2 G^2 r^4}{bD_0} \left[ \delta - t_c \frac{12}{41} \left(1 - e^{-\alpha_1^2 \frac{\delta}{t_c}}\right) \right]$$

$$= \frac{7}{48} \frac{\gamma^2 G^2 r^4}{bD_\parallel} \left[ \delta - \frac{r^2}{D_\parallel} \frac{12}{41} \left(1 - e^{-\alpha_1^2 \frac{D_\parallel \delta}{r^2}}\right) \right]. \quad [12]$$

To derive the previous expression, we utilised the following approximations:

$$\sum_{m=1}^{\infty} \frac{\alpha_m^2}{\alpha_m^6(\alpha_m^2-1)} \approx \frac{7}{48 \times 4},$$

$$\sum_{m=1}^{\infty} \frac{1}{\alpha_m^6(\alpha_m^2-1)} e^{-\alpha_m^2 \frac{\delta}{t_c}} \approx \frac{1}{\alpha_1^6(\alpha_1^2-1)} e^{-\alpha_1^2 \frac{\delta}{t_c}}, \quad [13]$$

$$\sum_{m=1}^{\infty} \frac{1}{\alpha_m^6(\alpha_m^2-1)} \approx \frac{1}{\alpha_1^6(\alpha_1^2-1)} \approx \frac{7}{48 \times 4} \times \frac{12}{41}.$$

Note that in Eq. [12], the radial diffusivity is non-negative $D_\perp(r) \geq 0$ for all radii satisfying the condition:

$$1 - \frac{12}{41} \frac{r^2}{D_0 \delta} \left(1 - e^{-\alpha_1^2 \frac{D_0 \delta}{r^2}}\right) \geq 0. \quad [14]$$

It can be shown that the inequality in Eq. [14] is valid for all values of $r$. For instance, in the limit of small $r$, $\delta \gg t_c$, Eq. [12] becomes equal to Eq. [11]. On the other hand, in the limit of large radii, the exponential term in Eq. [14] can be expanded using the Taylor series, and we obtain

$$1 - \frac{12}{41} \frac{r^2}{D_0 \delta}\left(1 - 1 + \alpha_1^2 \frac{D_0 \delta}{r^2}\right) = 1 - \frac{\alpha_1^2 12}{41} = 0.0078,$$  [15]

fulfilling the inequality $D_\perp(r) \geq 0$.

## 10.2 Appendix B

The effective radius can be determined using the following derivation:

$$\begin{aligned} \exp\left(-\frac{2TE\rho_2}{r_{eff-MRI-R}}\right) &\approx \frac{\int P(r)r^2 \exp\left(-\frac{2TE\rho_2}{r}\right)dr}{\int P(r)r^2 dr} \\ &\approx \frac{\int P(r)r^2 \left(1 - \frac{2TE\rho_2}{r}\right)dr}{\int P(r)r^2 dr}, \text{ for } \frac{2TE\rho_2}{r} \ll 1 \\ &= 1 - \frac{2TE\rho_2 \int P(r)r dr}{\int P(r)r^2 dr} \\ &\approx \exp\left(-\frac{2TE\rho_2 \int P(r)r dr}{\int P(r)r^2 dr}\right). \end{aligned}$$  [16]

Therefore,

$$r_{eff-MRI-R} \approx \frac{\int P(r)r^2 dr}{\int P(r)r dr} = \langle r^2 \rangle / \langle r \rangle.$$  [17]

## 10.3 Appendix C

**Figure A1** shows the spherical mean diffusion signal for various models, including the van Gelderen model (Eqs. [4] and [10]), the approximation for medium-pulse times (Eqs. [4] and [12]), the Neuman long-pulse limit (Eqs. [4] and [11]) and the first-order Taylor approximation of the Neuman model. Note that for fibre radii larger than 2.5 µm, the Neuman approximations deviate from the more accurate van Gelderen model. Conversely, the approximation for medium-pulse times produced accurate results.

Insert Figure A1 here (1 column)

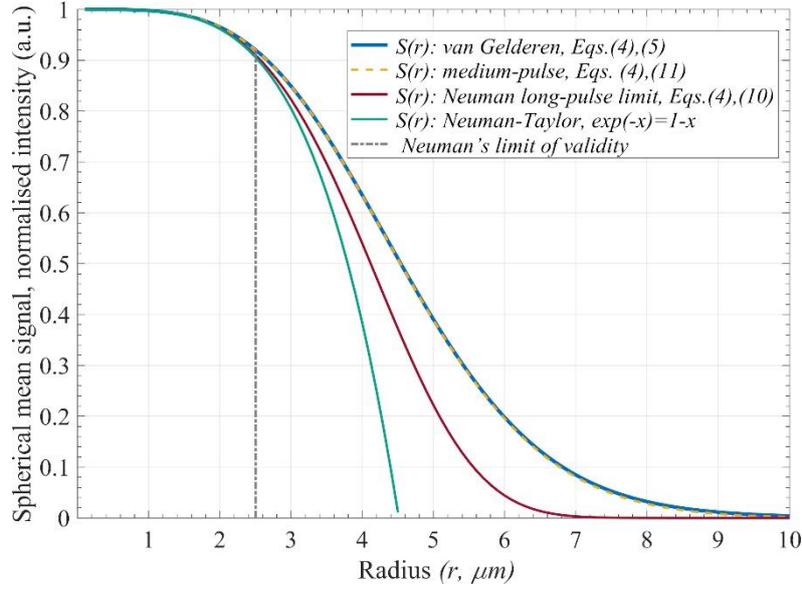

**Figure A1.** Spherical mean diffusion signal as a function of the radius for the acquisition sequence parameters employed in this study with $b$=10000 s/mm$^2$. Four models are displayed, including the van Gelderen model (Eqs. [4] and [10]), the medium-pulse approximation (Eqs. [4] and [12]), the Neuman approximation in the long-pulse limit (Eqs. [4] and [11]) and the first-order Taylor expansion of the Neuman model.

**Figure A2** displays the spherical mean diffusion and T$_2$ relaxation signals as a function of the fibre radius for the acquisition parameters employed in this study. Moreover, we plot the resolution limits for both normalised signals, defined as the minimum radius for which the signal deviates more than one noise standard deviation $\sigma$ compared to the signal generated for $r \rightarrow 0$. This definition considers that we cannot accurately detect signal decays smaller than $\sigma$. Note that the diffusion resolution limit is >1.4 µm, whereas the T$_2$-based resolution limit is much smaller, <0.2 µm. The T$_2$-based resolution limit for shorter *TE*s is even smaller (result not shown).

Insert Figure A2 here (1 column)

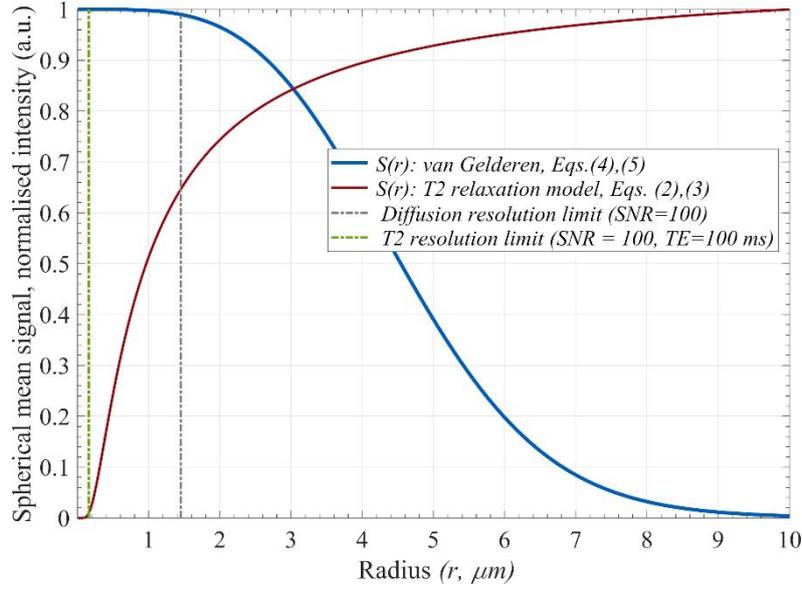

**Figure A2.** Spherical mean diffusion signal and $T_2$ relaxation signal as a function of the radius for the acquisition sequence parameters employed in this study. The diffusion signal was generated for $b$=10000 s/mm², and the $T_2$ relaxation signal was generated for $TE$=100 ms, using the parameters $\bar{\rho}_2$ =3.7 nm/ms and $T_2^b$ =3 s estimated in this study. The resolution limits are shown for the noise level $\sigma$ =1/100, i.e., SNR=100.

*10.4 Appendix D*

By considering Eqs. [1], [7] and [8] it is possible to demonstrate that neglecting the relaxation term in the spherical mean power-law approach leads to an effective radius estimate that corresponds to a distorted radius distribution $\tilde{P}(r)$,

$$\bar{S}_{\text{Diff}}(b, \tilde{r}_{eff-MRI-D}) \approx \frac{\int \tilde{P}(r) r^2 \bar{S}_{\text{Diff}}(b, r) dr}{\int P(r) r^2 dr}, \qquad [18]$$

where $\tilde{P}(r) = P(r)\exp\left(-TE/T_2^i(r)\right)$ is a distorted version of $P(r)$ due to the relaxation process not being modelled in a pure diffusion model. For a constant $TE$, the signal from the relaxation term $\exp\left(-TE/T_2^i(r)\right)$ is higher for larger $T_2^i$ times. As $T_2^i$ increases with $r$, the values of $\tilde{P}(r)$ for big radii are more inflated than those with small radii. Hence, this approximation leads to overestimating the effective radius calculated by the spherical mean power-law method.